\documentclass[10pt,letterpaper]{article}

\usepackage{opex3}

\begin{document}

\title{Demonstration of optically modulated
dispersion forces}

\author{F.~Chen,${}^1$ G.~L.~Klimchitskaya,${}^2$
V.~M.~Mostepanenko,${}^3$ and U.~Mohideen${}^1$
}
\address{${}^1$Department of Physics,
University of California,
Riverside, California 92521, USA.\\
${}^2$North-West Technical
University, St.Petersburg, 191065, Russia.\\
${}^3$Noncommercial Partnership ``Scientific Instruments'',
Moscow, 103905, Russia.}

\email{Umar.Mohideen@ucr.edu}

\begin{abstract}
We report the first experiment on
the optical modulation of dispersion forces through a change of the
carrier density in a Si membrane.
For this purpose a high-vacuum based atomic force microscope
and excitation
light pulses from an Ar laser are used.
The experimental results are compared with two
theoretical models.
The modulation of the dispersion force will find applications
in optomechanical micromachines.
\end{abstract}




\section{Introduction}

Dispersion forces \cite{1}, which is a generic name for
van der Waals and Casimir forces \cite{2},
 are of vital importance in diverse systems and phenomena,
such as membranes and layered structures \cite{4},
chemical and biological processes \cite{5},
carbon nanotubes \cite{7}, Bose-Einstein
condensation \cite{9},
noncontact atomic friction \cite{10}, nanoelectromechanical
devices \cite{13} and as a test for predictions of modern unification
theories [9--11].
Modern measurements of dispersion forces are reviewed in
\cite{26}.
The basic theory of dispersion forces was developed
by Lifshitz \cite{18}. However, the application of this theory to real
materials at nonzero temperature faces problems \cite{19}.
Here we first demonstrate the
optical modulation of dispersion forces
through a change in the carrier density by the absorption of
photons. For this purpose a high-vacuum based atomic force microscope (AFM)
is used to measure the modification in the force between a gold coated sphere
and a single-crystalline Si membrane. The excitation of the carriers in
Si is done with 514\,nm light pulses from an Ar laser.
Our experimental results
are in agreement with the Lifshitz theory if, in the absence of
excitation light, the model description of Si allows a finite
static dielectric permittivity. At the same time,
the model taking into
account the dc conductivity of high-resistivity
Si is excluded by our measurements. The provided experimental
results are topical for numerous applications of
dispersion forces ranging from biology, optomechanics, tribology,
condensed matter, atomic physics,
and to string theory.

\section{Experimental setup and measurement results}

Illumination with laser light
is an effective method to increase the carrier density of a
semiconductor up to values of order $10^{19}\,\mbox{cm}^{-3}$ required
to observe the modification of dispersion forces.
Our experimental setup used for
the optical modulation of the dispersion force is shown in
Fig.~1. A gold coated polystyrene sphere with a diameter
$2R=197.8\pm 0.3\,\mu$m is mounted on the tip of a 320$\,\mu$m conductive
AFM cantilever at a distance $z$ above a single-crystalline Si membrane in
a vacuum chamber. The thickness of gold coating on the sphere was measured
to be $82\pm 2\,$nm. An oil free vacuum with a pressure of around
$2\times 10^{-7}\,$Torr is used. The membrane is mounted on top of a piezo
capable of traveling a distance $z$ up to 6$\,\mu$m
between the test bodies. The complete movement of the piezo, $z_{\rm piezo}$,
was calibrated using a fiber optic interferometer. To extend and contract
the piezo, continuous triangular voltages with frequencies between
0.01--0.02\,Hz are applied to it. (Given that the experiment is done at
room temperature, applying of static voltages would lead to piezo creep
and loss of position sensitivity.)

\begin{figure}[bp]
\vspace*{-2.5cm}
\centering{
\includegraphics[width=14.5cm]{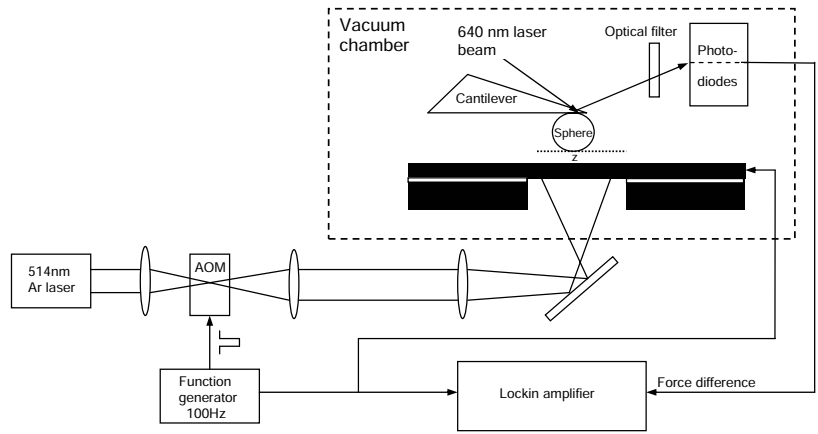}
}
\vspace*{-11cm}
\caption{Schematic of the experimental setup.
Light from a 514\,nm Ar laser is chopped into 5\,ms pulses and
irradiates a Si membrane leading to the modulation of the
dispersion force between the membrane and a sphere (see text
for further details).
}
\end{figure}
To modulate the dispersion force between the sphere and the membrane due to
the excitation of carriers, we illuminate the membrane with 514\,nm
pulses, obtained from a CW Ar ion laser. The light is
focused on the bottom surface of the membrane. The Gaussian width of
the focused beam was measured to be $w=0.23\pm 0.01\,$mm. The cantilever
flexes in response to the force changes. This deflection is monitored with
an additional 640\,nm laser beam reflected off the top of the cantilever
tip and leads to a difference signal between two photodiodes
(see Fig.~1). An optical filter
was used to prevent the interference
of the 514\,nm excitation light with the cantilever deflection signal.
The excitation laser light was modulated at a frequency of 100\,Hz
(5\,ms wide light pulses) using an Acousto-Optic-Modulator (AOM).
The AOM is triggered with a function generator. The same
function generator is
also used as a reference for the lockin amplifier, designed to measure
the difference of dispersion forces in response to the carrier excitation,
and for applying compensation voltages to the membrane (see below).

The illumination of the Si has to be done such that very little if any
light impinges on the sphere, as this would lead to a light induced force
from the photon pressure. As the Si membrane is illuminated from the
bottom, care should be taken that the fraction of light transmitted through
the membrane is negligibly small. Thus, the thickness of the membrane has
to be greater than  1$\,\mu$m which is the optical absorption depth of
Si at a wavelength of 514\,nm ($\omega=3.66\times 10^{15}\,$rad/s).
Calculations show that for our membrane of
about $4\,\mu$m thickness the force on a sphere due to photon pressure
varies from 2.7\% to 8.7\% of the difference of dispersion forces
to be measured when separation changes from 100 to 200\,nm.
Fabrication of the Si device with
a few micrometer thick single crystal
membrane of $<100>$ orientation (colored black in Fig.~1 with white
buried SiO${}_2$ layer) is necessary to
accomplish the experimental conditions.
It is achieved using a commercial Si grown on insulator wafer,
subject to mechanical polishing, RCA cleaning and TMAH etching
(details will be published elsewhere).
An ohmic contact is formed by a thin film of Au deposited on the edge of
Si device layer far away from its central part
followed by annealing at 673\,K for 10 min. The Si
membrane surface was cleaned with Nanostrip and then passivated by
dipping in 49\% HF for 10\,s. The passivated Si membrane was then
mounted on top of the piezo as described above.

The calibration of the setup, determination of the cantilever deflection
coefficient and the average separation on contact between the test bodies
are performed as in earlier experiments with metal and semiconductor
test bodies  [12,15--17]. For the determination of
the deflection coefficient, $m$, and the separation on contact, $z_0$,
we apply different dc voltages $V$ between 0.65 to --0.91\,V
to the membrane at large
separations from 1 to 5$\,\mu$m, where the dispersion force is negligible.
By fitting the experimental force-distance relation to the exact
theoretical expression $F_e=c(z)\left(V-V_0\right)^2$
for the electrostatic force, where $V_0$ is the residual potential difference,
$c(z)$ is a known function \cite{23}, we arrive
at $m=137.2\pm 0.6\,$nm per unit deflection signal $S_{\rm def}<0$ and
$z_0=97\pm 1\,$nm. Then the actual separations $z$ between the bottom of
the gold sphere and the Si plate are given by
$z=z_{\rm piezo}+mS_{\rm def} + z_0$. For the calibration of the deflection
signal and the determination of the residual potential between the two
surfaces, in addition to the dc voltages, a square voltage pulse of
amplitude from 1.2 to --0.6 V is also applied to the membrane \cite{23}.
By fitting the difference signal to the exact theoretical expression,
the calibration constant and the residual potential were obtained to be
$6.16\pm 0.04\,$nN per unit cantilever deflection signal and
${V}_0=-0.171\pm 0.002\,$V. All this was done in the same high
vacuum setup.

Next the carriers were excited in the Si membrane by 514\,nm laser pulses
and the difference in the total force (electric and dispersion) with and
without light
\begin{equation}\Delta F_{tot}(z)=c(z)\big[{(V^l-V_0^l)}^2-{(V-V_0)^2}\big]+
\Delta F_d(z).
\label{eq1}
\end{equation}
\noindent
is measured by the lockin amplifier with an integration
time constant of 100\,ms which corresponds to a bandwidth of 0.78\,Hz.
Here $\Delta F_d(z)=F_d^l(z)-F_d(z)$ is the difference in the dispersion
force where $F_d^l\,(F_d)$ is a force with (without) light.
$V_0^l$ ($V_0$)  is the still unknown
residual potential difference between the sphere and the membrane
during the bright (dark) phase of a laser pulse train
(they may be different from
the above-determined $V_0$ when no light is incident). $V^l$ ($V$) are
voltages applied to the membrane during the bright (dark) phase of
the laser pulse train.
By keeping $V=\mbox{const}$ and changing $V^l$, we measure the parabolic
dependence of $\Delta F_{\rm tot}$ as a function of $V^l$. The value of
$V^l$ where the parabola reaches a maximum [recall that $c(z)<0$] is $V_0^l$.
Then we keep $V^l=\mbox{const}$, change $V$ and measure the
parabolic dependence of
$\Delta F_{\rm tot}$ on $V$. The value of $V$ where this function reaches
a minimum is $V_0$. Both procedures were repeated at different separations
and the values $V_0^l=-0.303\pm 0.002\,$V and
$V_0=-0.225\pm 0.002\,$V were found to be independent of separation in
the range from 100 to 500\,nm reported below.

With these values of $V_0^l$ and $V_0$,
$\Delta F_d(z)$ at every
separation $z$ was determined from Eq.(\ref{eq1})
using the measured value of $\Delta F_{\rm tot}(z)$.
This was repeated with 41 pairs
of different applied voltages ($V_l,\,V$) and the mean value of
$\Delta F_{d}(z)$ was found.
Data were collected starting from contact at equal time intervals
corresponding to 3 points per 1\,nm. In Fig.~2 the
experimental data for the mean $\Delta F_d$ as a function of separation
varying from 100 to 500\,nm (1209 points)
are shown as dots. As is seen in Fig.~2, $\Delta F_d<0$, i.e., the
magnitude of the dispersion force with light is larger than without light
in line with physical intuition (recall that we follow the
definition of attractive forces as negative quantities).
The variance of the mean $\Delta F_d(z)$, $s(z)$, decreases from 0.16\,pN
at $z=100\,$nm to 0.11\,pN at $z\geq 250\,$nm.
Using Student's $t$-distribution with a number of degrees of freedom $f=40$
and choosing $\beta=0.95$ confidence, we obtain $p=(1+\beta)/2=0.975$ and
$t_p(f)=2.00$. This leads to the variation of the random error of
$\Delta F_d(z)$, equal to $s(z)t_p(f)$, from 0.34\,pN at $z=100\,$nm to
0.24\,pN at  $z\geq 250\,$nm. The systematic error in $\Delta F_d$ is
determined from the resolution error in data acquisition, from the calibration
error and from the total instrumental noise, and is equal to 0.09\,pN at all
separations. Thus, from statistical criterion \cite{24}, the total
experimental error at 95\% confidence is given by the random error.
As a result, the relative experimental error changes from 10 to 25\%
when the separation increases from 100 to 180\,nm. This allows us to conclude
that the modulation of dispersion force with light is demonstrated at a high
reliability and confidence.
The observed effect cannot be due to the mechanical motion
of the membrane. This is because vibrations due to heating
(in our case less than 1${}^{\circ}$C) would lead
to different force-distance relationship in disagreement with our
data in Fig.~2.
Within the separation range from 180 to 250\,nm
the experimental error is less than 46\%. At $z=360\,$nm it reaches 100\%.
Note that there was an early attempt \cite{25} to modify the dispersion
force between a glass lens and a Si plate with light. However, glass is
an insulator and therefore the electric forces such as due to work function
potential differences could not be controlled. This might also explain
that no force change occurred on illumination for small separations below
350\,nm where it should be most pronounced.

\section{Comparison with theory}

For comparison of our experimental results with theory we have calculated
the difference of dispersion forces $\Delta F_d(z)$ from the Lifshitz
formula. The calculations were done at the laboratory temperature
$T=300\,$K with the formula adapted for the configuration of
a sphere above a plate \cite{26}
using the dielectric permittivities of gold $\varepsilon^{Au}$
and Si $\varepsilon^{Si}$ along the imaginary frequency axis.
The $\varepsilon^{Au}(i\xi_j)$ at nonzero Matsubara frequencies
$\xi_j=2\pi k_BTj/\hbar$, where $k_B$ is the Boltzmann constant,
 was
found from the dispersion relation using the tabulated optical
data for the complex refractive index \cite{28}.
As was shown in \cite{16}, the use of tabulated data leads to
less than 0.5\% error in the Casimir force
[when the second body is a semiconductor,
the calculational results are independent of what is substituted for
$\varepsilon^{Au}(0)$]. In the absence of excitation light, the dielectric
permittivity of high-resisitivity Si, $\varepsilon^{Si}(i\xi)$, was also
found \cite{19} from tabulated optical data (dashed line in
Fig.~3). If we take into account the dc conductivity of high-resisitivity
Si, the dielectric permittivity (dotted line
in Fig.~3) is given by [15--17,20,21]
\begin{equation}
{\tilde{\varepsilon}}^{Si}(i\xi)={\varepsilon}^{Si}(i\xi)+
\big({\tilde{\omega}}_p^{(p)}\big)^2/{[\xi(\xi+\gamma^{(p)})\big]},
\label{eq2}
\end{equation}
\noindent
where ${\tilde{\omega}}_p^{(p)}$ and $\gamma^{(p)}$ are the plasma
frequency and the relaxation parameter for p-Si.
\begin{figure}[htb]
\vspace*{-6.8cm}
\centering{
\includegraphics[width=15cm]{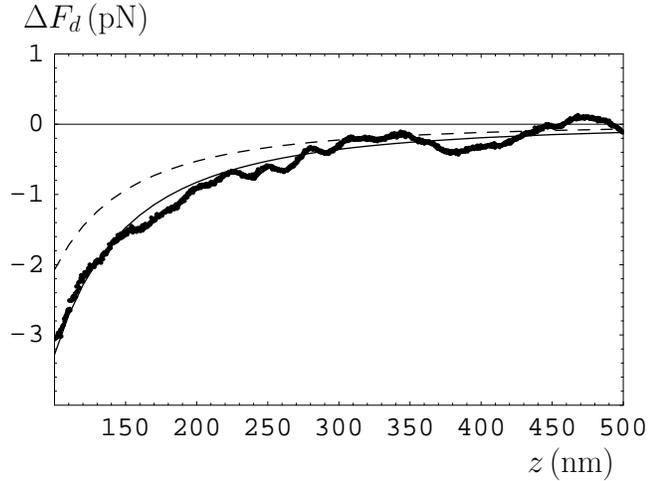}
}
\vspace*{-8.3cm}
\caption{ Differences of dispersion forces with laser
on and off.
Experimentally measured difference
data are shown as
dots. The solid line represents dispersion force difference
computed for Si with finite  static dielectric permittivity.
The force difference computed including the
dc conductivity of Si in the absence of laser light is
shown by the dashed line.
}
\end{figure}

In the presence of light, the equilibrium value of the carrier density is
rapidly established, during a period of time much shorter than the
duration of the laser pulse.
Therefore, we assume that there is an equilibrium
concentration of pairs (electrons and holes) when the light is
incident. The dielectric permittivity of Si in the
presence of laser radiation is commonly represented by the Drude
dielectric function which includes
free charge carriers  [15--17,20,21]
\begin{equation}
{{\varepsilon}}_l^{Si}(i\xi)={\varepsilon}^{Si}(i\xi)+
{\big({\omega}_p^{(e)}\big)^2}/{\big[\xi(\xi+\gamma^{(e)})\big]}+
{{\big(\omega}_p^{(p)}\big)^2}/{\big[\xi(\xi+\gamma^{(p)})\big]}.
\label{eq3}
\end{equation}
\noindent
It is shown as the solid line in Fig.~3.
Here the plasma frequencies $\omega_p^{(e,p)}$
and relaxation parameters $\gamma^{(e,p)}$ for electrons and
holes are introduced. The values of the relaxation
parameters and effective masses are \cite{29}
$\gamma^{(p)}\approx 5.0\times 10^{12}\,$rad/s,
$\gamma^{(e)}\approx 1.8\times 10^{13}\,$rad/s,
$m_p^{\ast}=0.2063m_e$, $m_e^{\ast}=0.2588m_e$.
From $\omega_p^{(e,p)}=\left[ne^2/(m_{e,p}^{\ast}\varepsilon_0)\right]^{1/2}$
with a charge carrier concentration \cite{28}
$\tilde{n}\approx 5\times 10^{14}\,\mbox{cm}^{-3}$ for Si of high
resistivity $\rho\approx 10\,\Omega\,$cm  we obtain
${\tilde{\omega}}_p^{(p)}\approx 2.8\times 10^{12}\,$rad/s.
In our experiment the uniform equilibrium
concentration of charge carriers induced by the laser radiation in
the region with a diameter  equal to the Gaussian width of the beam is
$n=4P_w^{\rm eff}\tau/(\hbar\omega d\pi w^2)$,
where $\tau$ is the excited carrier lifetime and
$P_w^{\rm eff}=3.4\pm 0.3\,$mW is the measured power absorbed for a
surface area $\pi w^2/4$.
The incident power is 13.7\,mW.
The lifetime $\tau=0.38\pm 0.03\,$ms was independently
measured using a non-invasive
optical pump-probe technique for the same membrane.
The Ar laser beam modulated at 100\,Hz
to produce 5\,ms wide square light pulses, as used in the Casimir force
measurement, was employed as the pump.
A CW beam with a 1\,mW power at a wavelength of 1300\,nm was used
as the probe. The change in the reflected
intensity of the probe beam in the presence and in the absence of
Ar laser
pulse was detected with a InGaAs photodiode. The reflected power of
the probe beam was monitored as a function of time in an oscilloscope
and found to be consistent with the change of carrier density.
(The details  will be reported elsewhere.)
This results in a concentration of charge
carriers induced by the incident light
$n=(2.0\pm 0.4)\times 10^{19}\,\mbox{cm}^{-3}$ and, as a consequence, in
$\omega_p^{(p)}=(5.6\pm 0.5)\times 10^{14}\,$rad/s,
$\omega_p^{(e)}=(5.0\pm 0.5)\times 10^{14}\,$rad/s.
A uniform carrier density in the membrane can be assumed,
because of
the long carrier diffusion lengths and the ability to obtain
almost defect free surfaces in silicon through hydrogen passivation
\cite{30}.

\begin{figure}[htbp]
\vspace*{-6.1cm}
\centering{
\includegraphics[width=14cm]{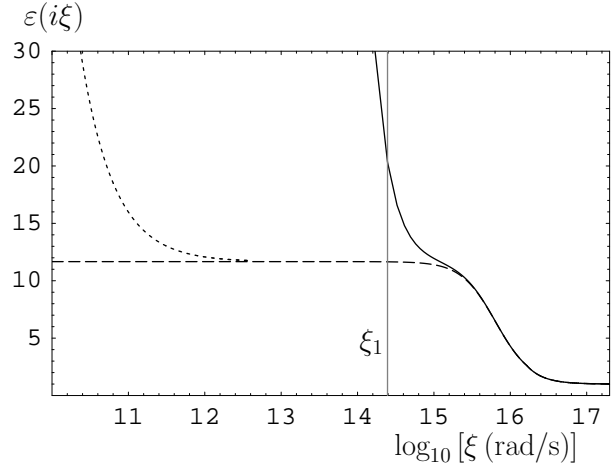}
}
\vspace*{-7.7cm}
\caption{ Dielectric permittivity of Si along the imaginary
frequency axis.
Solid line shows $\varepsilon_l^{Si}$ in the presence of laser light,
and the dashed line shows $\varepsilon^{Si}$ in the absence of light when
Si has a finite static permittivity.
${\tilde{\varepsilon}}^{Si}$ which includes the
dc conductivity in the absence of light is given by the dotted line.
$\xi_1$ is the first Matsubara frequency at $T=300\,$K.
}
\end{figure}
These values were used to calculate the theoretical force difference
using the Lifshitz formula for two models of the dielectric
permittivity of Si in the absence of laser light.
The topography of both surfaces was investigated using an AFM and
the effect of roughness \cite{31}
was taken into account as in \cite{23}.
It was found to be negligibly small.
The solid line in Fig.~2 represents the difference force
when in the absence of light the dielectric permittivity of Si is given
by $\varepsilon^{Si}(i\xi)$ with the value
$\varepsilon^{Si}(0)=11.66$.
In this case the transverse electric coefficient for Si
at zero Matsubara frequency is equal to zero as is true for any
dielectric, $r_{\rm TE}^{Si}(k_{\bot},0)=0$, where $k_{\bot}$
is the momentum component in the plane of membrane.
Thus, in the absense of laser light the obtained force does not
depend on the value of the same coefficient for Au,
$r_{\rm TE}^{Au}(k_{\bot},0)$, as only the product of both
coefficients enters the Lifshitz formula (recall that there are
different approaches in literature to the definition of
$r_{\rm TE}^{Au}(k_{\bot},0)$; see discussion in
Refs.~[9--11]). In the presence of light
$r_{\rm TE}^{Si}(k_{\bot},0)=0$ also holds true
due to the functional form of
the Drude model (\ref{eq3}). In both cases at zero frequency
only the transverse magnetic mode of the electromagnetic
field contributes to the result. For dielectric Si in the
absence of light
$r_{\rm TM}^{Si}(k_{\bot},0)=
[\varepsilon^{Si}(0)-1]/[\varepsilon^{Si}(0)+1]$.
For Si in the presence of light and for Au
$r_{\rm TM}^{Si}(k_{\bot},0)=r_{\rm TM}^{Au}(k_{\bot},0)=1$
holds.
The solid line in Fig.~2 is in excellent agreement with
data shown as dots.
If the permittivity of Si in the absence of light
is given by Eq.~(\ref{eq2}) (which includes
dc conductivity at frequencies much below the first Matsubara frequency
$\xi_1$ in Fig.~3),
the force difference calculated from the Lifshitz formula is
shown by the dashed line in Fig.~2 which disagrees with the experimental
dotted line.
In this case both in the absence and in the presence of light
$r_{\rm TE}^{Si}(k_{\bot},0)=0$ holds due to the properties of the Drude
dielectric function. Once again, at zero frequency only
the transverse magnetic mode contributes to the result.
Here, however, for Si in the absence of light
$r_{\rm TM}^{Si}(k_{\bot},0)=1$ holds.
Exactly this change in the magnitude
of the transverse magnetic reflection coefficient at zero frequency
leads to the deviation of the dashed line from the solid line in
Fig.~2. It can be considered as
somewhat surprising that the use of a more exact
dielectric permittivity (\ref{eq2}) instead of
$\varepsilon^{Si}(i\xi)$ leads to the discrepancy between
experiment and theory. This is, in fact, one more observation
that there are puzzles concerning the applicability of
the Lifshitz theory to real materials. In the case of
metals, the Drude description of conduction electrons in the
Casimir effect was excluded esperimentally in the series of
experiments [9--11].
In the case of metals, the deviation of the experimental results from
the Drude model approach is explained by the vanishing contribution
from the transverse electric mode at zero frequency.
The present experiment dealing with semiconductors
is not sensitive enough to detect this
effect. Here we report a novel effect due to
the difference in the contributions of the zero-frequency transverse
magnetic mode, which depends on whether or not
the dc conductivity of Si in the absence of light
is taken into account.
Note that in \cite{18} the dc conductivity of dielectrics
is not taken into account. The same is true for the recent paper
\cite{Cornell} on the thermal effects in the Casimir-Polder force.
This suggests that the theory of dispersion forces between
real materials requires further investigation.

The theoretical error in the computation
of force difference varies from 13.6 to 16.0\% when separation
increases from 100 to 200\,nm. Considering both the experimental and theoretical
errors, the model of a high-resistivity Si using Eq.(\ref{eq2})
is excluded by
our experiment within the separation region from 100 to 200\,nm at a 95\%
confidence.

\section{Conclusion}

In conclusion, we report the first experimental demonstration of
the modulation of dispersion forces through optical modification
of the carrier density of a Si membrane.
Such modulation can be used in the design and function of
optomechanical micromachines
such as micromirrors, nanotweezers and nanoscale actuators.

\section*{Acknowledgments}

The experimental section of the work was supported by the NSF Grant
No.~PHY0355092 and theory and analysis by DOE Grant No.~DE-FG02-04ER46131.
\end{document}